# Selecting competent referees to assess research projects proposals: a study of referees' registers[1]

**Abstract**

The selection of referees for evaluation of research projects under competitive financing appears particularly critical: the greater the competence of the referee concerning the core topic, the more their judgment is trustworthy and the more the financing of the best proposals is probable. The current work analyzes registers of experts used to select referees for the evaluation of research proposals in public programs with competitive funding. The work has the objective to present a methodology to verify the degree of "coverage" of the register compared to the spectrum of competencies necessary for the evaluation of such wide-ranging national programs; and to evaluate the level of scientific performance of the register's referees in the hard sciences, compared to the their national colleagues from the same fields.

---

# 1. Introduction

Currently, industrialized nations are increasing their efforts to improve efficiency in their domestic research systems, and one of the measures used for this purpose is the selective funding of research institutions and projects. Linked to the trend for selective funding is the ever wider occurrence of national research assessment exercises (VQR, 2011; ERA, 2010; RAE, 2008; etc.). These are primarily conducted through peer-review of the research products submitted by universities and other institutions, with reviews of the hard sciences sometimes supported by bibliometric methods. Peer-review alone has also always been used for the evaluation of actual research projects submitted for public funding, at international, national and regional levels.

It is frequently held that the just basis for research evaluation is for experts to review the works or project proposals of their colleagues. However, this process is exposed to inherent risks. Literature on the topic is very rich: for an extensive and excellent review we refer the reader to Wessely (1998). Briefly, one of the first points is that the exceptionally specialized nature of contemporary research makes it difficult to identify the most appropriate experts, and then, given their acceptance to serve as reviewers, that they succeed in recognizing the qualitative character of the product (a difficulty that increases with increasing specialization and sophistication in the research fields). Also, the rapidity of scientific advances can pose serious difficulties in contextualizing the quality of research outputs that in some cases have been produced a number of years previously. If a reviewer is called on to express a judgment on such a publication, will he/she be capable of discounting all the intervening scientific advances?

Peer-review is also clearly susceptible to built-in distortions from subjectivity in assessments (Moxham and Anderson, 1992; Horrobin, 1990). These can occur at several levels, including the early steps of selecting experts to carry out the assessments (Glantz and Bero, 1994). Further in the process, subjectivity in gauging output quality can result from real or potential conflicts of interest, for example from the experts' inclinations to give more positive evaluations to outputs from known and famous researchers than to those from younger and less established researchers. Bazeley (1998) conducted an empirical analysis of applications submitted to the Australian Research Council Large Grants Scheme to determine the influence of biographical and academic track-record on ratings by independent assessors, and showed that distortional effects in the judgments are present and can be traced to the renown of the proposing scientists. Langfeldt (2006) subsequently conducted an analysis of the characteristics and challenges facing peer-review from a variety of perspectives, including an examination of how reputation and social networking influence peer-review processes. The risk of such effects increases when lists of national referees are limited to or concentrated on scientists from within the same country: in these contexts it is quite possible that the evaluation process would not be totally "blinded", and that there will be further distortion in results. Further, the higher the ratio of products or projects to number of referees in the register, the higher the level of possible distortion. The ethical conduct of the referees here plays an important role, especially if the outcome of the evaluation is related to the interests of the referee or of his/her organization. Finally, peer-review methodology does not have universal consistency, since mechanisms for assigning merit are established autonomously by the various evaluation panels and/or individual reviewers, thus exposing the comparative methodology to further potential distortions. In summary, "bias in peer-review, whether intentional or inadvertent, is widely recognized as a confounding factor in efforts to judge the quality of research" (Pendlebury, 2009).

While some distortions are inherent to peer-review methodology, with little or no resolution possible, others, in particular those related to the selection of the referees, can be definitely addressed. To reduce risks in selecting referees and avoid distortions that could seriously undermine the effectiveness of an evaluation, the selection process must adhere to two fundamentals: 1) the closest matching possible between the competencies of referees and the subjects of the products or projects submitted; 2) excellence of the referees in their fields of research.

Supranational institutions such as the European Union, along with many national and regional governments and international journals resort to prepared registers of experts to identify the most appropriate referees for evaluation of research products and projects whether for ex-ante, in progress or ex-post evaluation. The decision to access a prepared register implies the risk of drawing on an assemblage that is not always adequate, in terms of the scientific coverage and the quality of the experts. The risk becomes greater with lesser size and internationalization of the register, and the problem is currently exacerbated by a decline in reviewers' willingness to participate in peer review exercises (ESF, 2011). It is not surprising then that a recent survey of the European Science Foundation (ESF, 2011) clearly identified the need and desire for a common European reviewer database. The financial incentives involved in refereeing are in fact generally low, and the referee's natural ethical conscience to serve his/her scientific community may naturally fade away. In some cases the opportunity to represent the interests of one's own country or institution may act as a more powerful incentive.

Assessment of the adequacy of referees' registers and the efficiency in matching referees with manuscripts or projects to be evaluated is not an easy task. In fact, for privacy reasons the general public rarely has access to the referee registers, not to mention the identity of the referees who have actually evaluated a particular paper or a project. This probably explains our unsuccessful literature search for previous studies on the subject. In Italy, the selection of experts for evaluation of public-call competitive research projects draws on a specific national register, purposely created by the Ministry of Education, Universities and Research (MIUR) in 2001. The MIUR developed the register using a "bottom up" procedure: a public call and selection where all applicants presented their requests for inclusion in the list, accompanied by information they felt would be useful for identifying their expertise in specifically named fields. Italy's regional governments also use the list, often drawing only from those referees that reside outside their relative territories. Through a study commissioned to the authors by a regional administration, we were provided access to the identity of the referees who actually evaluated research projects and authorized to carry out the investigation reported in this paper. In this work then, we propose a methodology to verify the level of adequacy of national registers, analyzing: i) the level of "coverage" of the register with respect to the complete spectrum of disciplines; ii) the level of scientific performance of the referees with respect to the subjects being evaluated, namely the authors of the proposed projects and works. We will then consider a specific case of use of the register at the regional level, to analyze the adequacy of the referees, as selected, with respect to the subjects they evaluate.

The aim of the paper is not to remark on any weaknesses particular to the Italian register, rather to present the methodology that we have adopted, which can also be applied in other national contexts to verify the adequacy of domestic registers with respect to the objectives for which they were developed.

In the next section we describe the pertinent characteristics of the Italian higher education funding system. In Section 3 we respond to the first research question, presenting data concerning the composition of the register. In Section 4 we respond to the second research question, first introducing the specific method of bibliometric analysis, and then the principle results from its application. Section 5 presents a field analysis concerning a call for proposals conducted in 2008 in the Region of Sardinia. The closing section provides the authors' comments on the analytical results and their implications.

## 2. The Italian higher education funding system

In Italy, the MIUR recognizes a total of 95 universities as having the authority to issue legally-recognized degrees. With only rare exceptions, these are public institutions that are largely financed through non-competitive government allocation. Until 2009, the core funding was input oriented, i.e. distributed to universities in a manner intended to equally satisfy the needs and resources of each and all, in function of their size and activities. The share of this core funding relative to total university income is now being reduced, descending from 61.5% in 2001 to 56.0% in 2009 (MIUR, 2011). It was only following the first national research evaluation exercise (VTR, 2006), conducted between 2004 and 2006, that a minimal share, equivalent to 7% of MIUR financing[2], was attributed in function of the assessment of research and teaching quality. Further financing from the MIUR for research projects on a competitive basis represents an additional 9% of overall income, with the major part of this distributed under two programs: PRIN, or "Research Projects of Relevant National Interest", and FIRB, or "Fund for Incentive in Basic Research" (inactive since 2003).In these programs, merit classification lists are based on peer judgments of proposals submitted by academic candidates. The referees are selected from a list specifically prepared by the MIUR: the following section describes and examines the composition of the list.

## 3. Degree of coverage for the MIUR register

In 2001, the MIUR developed a list of experts that it draws on to evaluate the technical-scientific aspects of research projects proposed under its PRIN and FIRB funding programs. The referees' register was developed by a public selection procedure[3] in which candidates presented their demands for participation, attaching all information which might assist in ascertaining their competencies in specific fields. As a prerequisite, the candidates also had to qualify as belonging to one of three categories of scientists:
 A) an associate or full professor;
 B) a research director or senior researcher in a public research institution;

---

[2] Government intention is that this share will gradually increase to 30%.
[3] A recent MIUR ministerial decree provided for updating the register, as seen at http://attiministeriali.miur.it/anno-2010/aprile/dm-01042010-n-79.aspx, last accessed on September 19, 2012.

C) an individual with at least 10 years of documented technical-scientific experience in a public or private research organization.

Candidates also had to indicate one or more areas of competence selected from two classification systems:

- the National Institute of Statistics classification of economic activities (ATECO-ISTAT)[4],
- the MIUR classification of scientific disciplines in the Italian university system[5].

The MIUR system spells out a total of 370 specific "scientific disciplinary sectors" (SDSs)[6], which are grouped in 14 disciplines called "university disciplinary areas" (UDAs).

A specific ministerial commission was then responsible for evaluating the candidates, drawing on the education records and other documentation to assess the competencies they declared. As an example, Table 1 shows the competencies that were declared and ascertained as present for an expert in category A of the register.

[Table 1]

The register lists a total of 1,545 research experts: 800 in category A, 304 in B, 307 in C, with no category indicated for the remaining 134. From this list we identified 978 Italian academics, with most of these being associate and full professors (category A) and all the remainder being assistant professors (category C).

For the purpose of this study we limit our attention to the competencies differentiated under the MIUR classification scheme, meaning the UDAs and the more specific SDSs. For reasons of significance, which we explain below, we then limit the analysis to the 205 SDSs in the hard sciences[7].This is only a modest limitation, given that these hard science SDSs include 1492 of the MIUR experts (956 academics), which is 97% of total. Table 2 shows the SDSs and experts, by UDA. Note that the individual experts can present candidatures in more than one SDS. This means that column three and four data do not coincide with the bottom line totals.

[Table 2]

It is immediately obvious from the data that there is a non-uniform distribution of experts: in the Industrial and information engineering SDSs there are 774 experts, or 38% of the overall total. Limiting observation only to academics, the incidence for this UDA rises to 42% of total. Earth science is at the opposite extreme: the SDSs of this

---

[4] http://www.oldstarnet.unioncamere.it/intranet/Area-Terri/Toscana/Documenti/Classifica/Ateco-91.xls_cvt.htm, last accessed on September 19, 2012

[5] http://attiministeriali.miur.it/UserFiles/115.htm, last accessed on September 19, 2012

[6] As an example, we quote the SDS description for FIS/03-MATERIALS PHSYICS: "The sector includes the competencies necessary for dealing with theory and experimentation in the state of atomic and molecular aggregates, as well as competencies suited to dealing with properties of propagation and interaction of photons in fields and with material. Competencies in this sector also concern research in fields of atomic and molecular physics, liquid and solid states, semiconductors and metallic element composites, dilute and plasma states, as well as photonics, optics, optical electronics and quantum electronics".

[7] Belonging to the following UDAs: Mathematics and computer science, physics, chemistry, earth science, biology, medicine, agriculture and veterinary science, civil engineering, industrial and information engineering

UDA account for only 54 persons (3% of total) in the list of experts, with 21 being academics.

Prior to an in-depth analysis of the level of disciplinary coverage, we already note that for 18 of the 205 SDSs examined there are no experts included in the list (Table 2, column 5). Any potential project centered on themes typical of these SDSs could not be evaluated by adequately competent referees. The problem is particularly serious in medicine, where there is an absence of experts for 11 SDSs out of the 50 for the UDA. On the other hand, there is at least one expert for every SDS in Industrial and information engineering, Chemistry, Physics and Mathematics and computer science. In reality the presence of a single expert in an SDS still represents a problem, since review of a project must normally be entrusted to at least two referees. The last column of Table 2 quantifies the problem, showing that 25 SDSs out of 205 are "covered" by a single expert. Potential projects with content situated in these SDSs can only be partially evaluated, not having a second competent judgment from other than the lone expert present in the register. Once again, the Medicine UDA is most problematic, with 16 SDSs (32% of 50 total) covered by a single expert. Summarizing these observations, for 43 of a total 205 SDSs, a potential project would not receive any competent evaluation or would receive only a "partial" evaluation.

Still dealing with the distribution of experts among the SDSs, one of the relevant characteristics of the register is the level of coverage (again in terms of competencies) compared to what could be the true needs of an agency, a ministry, or of any organization that intends to identify experts to evaluate research projects. If we assume a reference context similar to the MIUR financed programs such as the PRIN or FIRB, where eligibility is limited to the university research staff[8], we could expect that the desirable levels of disciplinary coverage in the register would more or less align with staff distribution per UDA (and at a secondary level, per SDS). Table 3, third column, shows the actual distribution of staff per UDA: the values in parentheses show the percentage incidence of each UDA in total national academic staff and can thus be likened to the incidence of demand for evaluation for each UDA. The second column shows the distribution of academic experts registered in the list. For this specific table, the SDS (and thus UDA) associated with each expert is the one of his/her official faculty role (as won in national competition for tenure) and not the one(s) indicated in his/her application to the register. In effect, as seen in Table 1, the academic experts are accredited in varying numbers of SDSs (average 4.2), but it is reasonable to hold that the SDS of official membership is the one where they will show greatest competence in evaluation. The ratio between values in the second and third columns provides a concentration index that expresses the adequacy of the register in terms of disciplinary coverage. We assume that researchers of all the UDAs will participate with equal enthusiasm in calls for financing proposals and thus create demands for evaluation that are proportional to the size of the UDAs. Thus, when last column values are higher or lower than one, the table indicates that the register actually lists a greater or lesser number of experts than is desirable. A value for a UDA that is far from one shows an imbalance of the experts in the register in relation to national research staff.

[Table 3]

---

[8] FIRB calls for proposals are open to all not-for-profit research organizations.

It is immediately obvious that there is a lack of balance in the composition of the register in terms of its representativeness of the UDAs: 46% of the academics registered are faculty in Industrial and information engineering. This presence is over three times (3.30) the incidence in the national academic staff. At the opposite extreme of concentration are Medicine (0.29) and Earth science (0.37). Medicine is actually the largest national UDA, with over 12,000 scientists, equal to 31% of research staff employed in the hard sciences. Yet in the MIUR register there are only 85 experts that belong to this UDA (9% of total). Earth science is a smaller UDA (4% of Italian research staff in the hard sciences) but still with only 13 experts registered in the list. Biology (0.57) and Mathematics and computer science (0.58) also show values for concentration index that are less than one. Again, these are cases where the representativeness of experts appears totally inadequate to deal with the potential demand for evaluation of financing proposals, for which the register was conceived.

There is also another critical problem in the disciplinary coverage: 37 SDSs are not actually covered by academic experts officially belonging to them, but by scientists from other SDSs. It appears that the MIUR may not adequately check situations where professors that belong to one SDS propose themselves as experts for another one, and thus that there is a form of "cross-colonization" of some fields. By checking the data we can identify two quite critical examples:

- for SDS ING-IND/05, only six professors, all from other SDSs, applied and were accepted; of these, three are officially classified in SDS ING-IND/08;
- for SDS MED/06 there are twelve academic referees, all from other SDSs, including six from MED/04.

For examples such as these two SDSs, there is the real possibility that a small group of referees from other SDSs would dominate the evaluation of proposed projects and articles.

## 4. Research performance of experts on the MIUR register

The preceding section shows some critical concerns with the register used by the MIUR for selection of experts to evaluate competitively-funded research projects. These concerns mainly regard the degree of coverage relative to the spectrum of disciplines concerned by the calls for proposals. In this section we now attempt to evaluate the adequacy of the register in terms of level of scientific activity of the experts included. The analysis is again limited to the 956 academic experts active in the hard sciences, this time using a bibliometric methodology (described in the next subsection) that compares their research productivity with that of their national colleagues in the same SDS but not included in the register. A more informative comparison to an international standard, however desirable, is not feasible yet, because of the difficulty of disambiguating foreign authorships and classifying foreign scientists by fields of research.

**4.1 Methodology**

Research activity is a production process in which the inputs consist of human, tangible (scientific instruments, materials, etc.) and intangible (accumulated knowledge, social networks, etc.) resources, and where outputs have a complex character of both tangible nature (publications, patents, conference presentations, databases, protocols, etc.) and intangible nature (tacit knowledge, consulting activity, etc.). The new-knowledge production function therefore has a multi-input and multi-output character. The principal efficiency indicator of any production system is labor productivity. To calculate it we need to adopt a few simplifications and assumptions. In the hard sciences, including life sciences, the prevalent form of codification of research output is publication in scientific journals. As a proxy of total output in this work we consider only publications (articles, article reviews and proceeding papers) indexed in the WoS. For the hard sciences, the literature gives ample justification for the choice of the bibliometric approach, reasoning that scientific publications are a good proxy of overall research output (Moed et al., 2004). The other forms of output which we neglect are often followed by publications that describe their content in the scientific arena, so the analysis of publications alone actually avoids a potential double counting.

When measuring labor productivity, if there are differences in the production factors available to each scientist then one should normalize by them. Unfortunately, relevant data are not available at individual level in Italy. The first assumption then is that resources available to professors within the same field of observation are the same. The second assumption is that the hours devoted to research are more or less the same for all professors. In Italy the above assumptions are acceptable because in the period of observation, core government funding was input oriented and distributed to satisfy the resource needs of each and every university in function of their size and activities. Furthermore, the hours that each professor has to devote to teaching are established by national regulations and are the same for all. Because research projects frequently involve a team of researchers, which shows in co-authorship of publications, productivity measures then need to account for the fractional contributions of scientists to their outputs. Furthermore, because the intensity of publications varies across fields (Abramo et al., 2008), in order to avoid distortions in productivity rankings, one must compare researchers within the same field. A prerequisite of any distortion-free research performance assessment is thus a classification of each researcher in one and only one field. In fact, in the Italian university system all professors are classified in one field. This feature of the Italian higher education system is unique in the world. In the hard sciences, there are 205 such fields (named scientific disciplinary sectors, SDSs[9]), grouped into nine disciplines (named university disciplinary areas, UDAs[10]).

A very gross way to calculate the average yearly labor research productivity is to simply measure the weighted fractional count of publications per researcher in the period of observation and divide it by the full-time equivalent of work in the period. A more sophisticated way to calculate productivity recognizes the fact that publications, embedding the new knowledge produced, have different values. Their value depends on their impact on scientific advancements. As proxy of impact, bibliometricians adopt the number of citations for the researchers' publications, notwithstanding the possible distortions inherent in this indicator (Glänzel, 2008).

---

[9] The complete list is accessible on http://attiministeriali.miur.it/UserFiles/115.htm, last accessed on September 19, 2012.
[10] Mathematics and computer sciences; physics; chemistry; earth sciences; biology; medicine; agricultural and veterinary sciences; civil engineering; industrial and information engineering.

However, comparing researchers' performance by field and academic rank is not enough to avoid distortions in rankings. In fact citation behavior also varies across fields, and it has been shown (Abramo and D'Angelo, 2011) that it is not unlikely that researchers belonging to a particular scientific field may also publish outside that field (a typical example is statisticians, who may apply their theory to medicine, physics, social sciences, etc.). For this reason we standardize the citations for each publication accumulated at June 30, 2009 with respect to the median[11] for the distribution of citations for all the Italian publications of the same year and the same subject category[12].

In formulae, the average yearly productivity at the individual level, named as Fractional Scientific Strength (FSS), is the following.

$$FSS = \frac{1}{t}\sum_{i=1}^{n} \frac{c_i}{Me_i} * \frac{1}{s_i}$$

Where:

t = number of years of work of the researcher in the period of observation
n = number of publications of the researcher in the period of observation.
$c_i$ = citations received by publication *i*;
$Me_i$ = median of the distribution of citations received for all Italian cited-only publications of the same year and subject category of publication *i*;
$s_i$ = co-authors of publication *i*

In the life sciences, widespread practice is for the authors to indicate the various contributions to the published research by the positioning of the names in the authors list. For the life sciences then, when the number of co-authors is above two, different weights are given to each co-author according to his/her position in the list and the character of the co-authorship (intra-mural or extra-mural). If first and last authors belong to the same university, 40% of citations are attributed to each of them; the remaining 20% are divided among all other authors. If the first two and last two authors belong to different universities, 30% of citations are attributed to first and last authors; 15% of citations are attributed to second and last author but one; the remaining 10% are divided among all others[13].

The period of observation for the publications is the 2004-2008 quinquennium: five years seems sufficient time for a robust evaluation of the research performance of individual academics. Also, we note that the register was first completed at the end of 2002, and that therefore its first use can only have been for funding programs starting in 2003.

The data used are taken from the ORP (Observatory of Italian Public Research[14]) a database that the authors derive from the Thomson Reuters Web of Science (WoS).

---

[11] As frequently observed in literature (Lundberg, 2007), standardization of citations with respect to median value rather than to the average is justified by the fact that distribution of citations is highly skewed in almost all disciplines.

[12] The subject category of a publication corresponds to that of the journal where it is published. For publications in multidisciplinary journals the scaling factor is calculated as a weighted average of the standardized values for each subject category.

[13] The weighting values were assigned following advice from Italian professors in the life sciences. The values could be changed to suit different practices in other national contexts.

[14] www.orp.researchvalue.it last accessed on September 19, 2012.

Beginning from the raw data indexed in the WoS, then applying a complex algorithm for reconciliation of the authors' affiliation and disambiguation of the true identity of the authors, each publication (article, article review, conference proceedings) is attributed to the Italian university scientists that produced it, with an error of less than 5% (D'Angelo et al., 2011). Each scientist is thus compared to all other Italian colleagues in the same SDS, and a percentile ranking is provided.

### 4.2 Results and analysis

Applying the above bibliometric methodology, we now wish to verify the level of scientific activity of the 956 academic scientists listed in the MIUR register and particularly to understand if their research performance is superior to that of national colleagues in the same SDS. Table 4 synthesizes the results of the analysis. We see that 5% of the referees on the list (second column, last line) show nil performance, either for lack of publications over the five years under observation or for lack of citations to works they may have published. Of the 908 actives, 322 (34%) show performance below the national median and 190 of these actually place in the last quartile. On the other hand, there are 350 listed scientists (37% of total) with "top class" performance, meaning that they fall in the first quartile at national level (last column, last line of table).

Analysis at the UDA level offers points for reflection, with certain UDAs presenting further critical concerns. In Civil engineering, 43% of the referees were inactive over the quinquennium and only 30% of the referees placed above the national median, the lowest percentage among all UDAs. In Agriculture and veterinary science, 14% of the referees were inactive and a further 31% place under the national median. Industrial and information engineering also registers a very high percentage of referees with performance below national median (43%), although in this UDA there are very few inactive referees (9, or 2% of total). Among the other UDAs, Biology seems to show a significant concentration of referees of high scientific profile: of 80 total referees only one is inactive, while 53% of the referees place in the first national quartile for performance. There are also above 50% top class referees in Medicine and Chemistry.

[Table 4]

The observations that should raise concern are the instances of significant numbers of inactive referees or significant numbers placing in the last national quartile. The entity of these combined subsets is obtained by summing the values of the second and third columns in Table 4. This combination represents 25% of the overall total and varies from a minimum of 7% in Biology to a maximum of 66% in Civil engineering. In substance, a quarter of the scientists in the register show nil or very limited scientific activity. For Civil engineering the situation concerns every two out of three referees.

Clearly the analysis by UDA does not exhaust the subject, since this is a highly aggregate level of classification, and a deeper examination at the SDS level would be interesting. Table 5 provides a simple count of the SDSs where the average level of referee performance is particularly high (top 20%) or, in contrast, completely insufficient (below national median). The values in column three show that only 28 out of the 187 SDSs covered by the register (15% of total) have average performance of referees that is top class. Mathematics and computer science is notable for the presence of three SDSs out of the total 10 with a high concentration of very active referees. On the other hand, we see that in 45 SDSs (24% of a total 187) the average value for performance by referees is insufficient: in these SDSs, there is a risk that evaluation of projects would be entrusted to experts with an inadequate scientific profile. The risk is yet higher in Industrial and information engineering, where the problem affects fully half of the SDSs (21 of total 42).

[Table 5]

**5. Call for proposals for research projects in the Region of Sardinia**

As shown in the previous section, resorting to a prepared register of referees implies the risk of drawing on an assemblage that is inadequate in both scientific coverage and quality of experts. We expect this risk to increase with reduction of size and internationalization of the register. We investigate the hypothesis through a field analysis of a specific call for proposals launched by an Italian regional administration where, for evaluation of proposals, the administration drew on the MIUR register to extract non resident experts.

Italy is subdivided in 20 regions with a certain level of administrative autonomy, including in policy for support of research activity carried out in regionally based organizations. In a 2007 law[15], the Region of Sardinia defined a specific intervention for promotion and support of the regional research system, and in 2008, issued a call for proposals[16] in fundamental or basic research, open to two categories:
- professors and researchers from the two universities in Sardinia (University of Cagliari, University of Sassari);
- researchers from research institutions, hospital and health agencies in the region.

There were a total of 467 projects presented, with 414 from the university category and 53 from scientists in the second category. For evaluation of the projects the region selected 75 referees from the MIUR register, of which 69 were academics and 6 were non-academics, all resident outside Sardinia.

Concentrating on the subpopulation belonging to the hard sciences, there were 382 subjects (327 proponents[17], 70% of the total, and 55 referees, 73% of total). For these, we conducted an analysis according to the methodology described above (Section 4.1). Table 6 presents the analysis of the selected referees' coverage with respect to the applicants' fields of research. The distribution of the applicants and referees by UDA is shown in columns two and three, Table 6. The ratio of the overall numerosity of referees to applicants is 1 to 6 (55/327). Physics and Medicine register the lowest values of this ratio: one referee for every 8 applicants. The applicants fall in a total of 130 SDSs, of which 44 are "covered" by referees (last line, Table 6): the difference (86) indicates the number of SDSs not covered by referees, in terms of the necessary competencies to evaluate all projects. The projects proposed by the applicants in these 86 SDSs (191 projects, 58% of total) were evaluated by referees working in SDSs different from those of the applicant. The last four columns of Table 6 present the data concerning this situation for the individual UDAs. The most notable case is definitely Industrial and information engineering, where 83% of the applicants were evaluated by referees from other SDSs.

---

[15] Regional Law 7/08/2007, no. 7 – *Promozione della ricerca scientifica e dell'innovazione tecnologica in Sardegna* (Promotion of scientific research and technological innovation in Sardinia), http://www.regione.sardegna.it/j/v/80?s=53788&v=2&c=3311&t=1 last accessed on September 19, 2012.
[16] Regional Decree 19/12/2008, no. 72/1, http://www.regione.sardegna.it/documenti/1_73_20090227133530.pdf, last accessed on September 19, 2012.
[17] The "proponent" is the project coordinator.

[Table 6]

Bibliometric analysis of the referees' scientific activity (Table 7) reveals a number of critical concerns: 22 referees (40% of total) have performance below the national median ($50^{th}$ percentile) and of these, eight are actually inactive over the five year period, showing nil performance[18]. Analysis by individual UDA shows notable differentiation: the most striking case is Agriculture and veterinary science, where five out of the eight referees show a performance lower than the national median, and among these, three are inactive over the five years observed. On the other hand, all the five referees in Civil engineering and Physics show performance above the national median.

[Table 7]

## 6. Conclusions

Many institutions resort to prepared registers of experts, drawing on them to select referees for evaluation of research projects or products, for national research assessment exercises and in support of supranational, national and regional funding calls. This decision implies the risk of drawing on an assemblage that is not always adequate in terms of degree of scientific coverage and quality of the referees. The risk becomes greater with lesser size and internationalization of the register.

In this work we have applied an assessment methodology and provided empirical evidence of the problem, analyzing the composition of a register created in 2001 by the Italian Ministry of Education, Universities and Research (MIUR): it serves for the selection of referees whose judgment guides assignment of funds under important national programs. The register is prepared by a "bottom-up" public selection procedure, in which interested single candidates prepare their requests for insertion in the list, with a self-declaration of their competencies in specified fields.

A first analysis shows that the register's level of coverage with respect to the complete spectrum of research fields is inadequate. Limiting the analysis only to the referees from universities, we observe that their distribution among fields is not at all uniform: Industrial and information engineering are "oversized", while for Medicine, Earth science, Biology, Mathematics and computer science, the referees seem too few compared to the potential demand for evaluation, estimated by the distribution of total national research staff. The register is without any referees in 18 of the 205 scientific SDSs identified by MIUR's own classification system; in another 25 SDSs there is only one referee: potential projects in these 43 SDSs (equal to 21% of total) would not be evaluated in adequate fashion, either because there are no referees or because there is no opportunity for a second competent judgment. The situation is different among the disciplines: this type of problem is particularly serious in Medicine, where it develops in 27 of its 50 total SDSs.

---

[18] Either because they didn't publish or didn't receive any citations.

A further bibliometric analysis also shows that the scientific profile of the listed referees is often inadequate relative to the potential subjects to be evaluated. We see that 5% of academic referees are scientifically inactive, either not producing any publications or not receiving any citations for production during the 2004-2008 period. Another 34% show bibliometric performance below the national median, and 20% place in the last quartile. Again, the analysis by discipline shows differentiation: in Civil engineering, Industrial and information engineering and Agriculture and veterinary science, the risk of depending on a referee with inadequate scientific profile is significantly high.

The analysis of a specific call for proposals by the Region of Sardinia confirms the entity of these risks. For this initiative, the regional administration drew on the national register to select from the referees listed by the MIUR, choosing only non-residents of Sardinia to carry out the evaluation of the proposed research projects.

Even though confidentiality rules make it impossible to precisely check the alignments of referee and applicant competencies, we can estimate that 58% of applicants were evaluated by referees with competencies foreign to their specific SDS. Further, 15% of the referees involved were scientifically inactive and 40% had a bibliometric performance that was under the national median.

All this seems to suggest a reconsideration of the choice to resort to registers that are prepared with bottom-up logic. Bibliometrics is now widely used in the area of research assessment, and produces classifications that make it possible (at least in the hard sciences) to have a top-down type of process for the identification of experts to use in peer-review of research proposals. Through appropriate elaborations of bibliometric data, it is possible to have decision support systems for the objective and unequivocal identification of who does what and how well, in every field of knowledge. When referees are to be chosen in the national context, a top-down process could be applied, using the same database as in the bibliometric analysis of this work. The distinctive characteristic of the ORP, different from other international data bases such as WoS itself, is the primary function for which it was conceived: to identify, through classification of productivity and quality of scientific output, the best competencies in the nation, in whatever field of research interests the user. An open query for any subject results in all the authors (and institutions), in order of productivity and quality, who have written articles on that subject. However, when referees are to be chosen in the international context, databases such as WoS and Scopus would serve the same purpose. The procedure to identify highly cited referees in the field of interest would be fairly complex, however it would be simple to search for referees with very high numbers of publications in the field. Using the WoS, for example, it would be quite easy to set a search for specific keywords relating to the main topics of any project. The user can refine the query to specific subject categories if the keywords are generic and likely to draw articles from disciplinary areas unrelated to the project topic. The search can also be limited to a specific time window in order to pick the most recent literature. The results of these queries can be easily analyzed through functions already available in the WoS, such as "analyze results", with setting "rank the records by author". Similar procedures can also be readily conducted on Scopus. These systems are certainly more effective than any register developed through a public call and based on self-declared data, and are adequate for any current evaluation need without limit on size. Further, the fact that the commercial services constantly update the source bibliometric databases offers obvious advantages in cost and time, relative to upgrading lists by the "public call" approach. In the context of growing attention to increased efficiency in research systems, decision makers should seriously consider adopting these more effective, methods to support their choices in the very awarding of research financing. Editors of scientific journals could also consider the same methods for better matching to the content of proposed publications when they select experts to serve as reviewing referees.

| Competence system | Code | Description |
|---|---|---|
| ATECO ISTAT '91 | 34 | Manufacturing: vehicles, trailers and semi-trailers |
| | 35 | Manufacturing: other transport |
| | 40 | Production of electrical energy, gas, steam or hot water |
| | 29 | Manufacturing: machines and mechanical equipment, including installation, set up, repair, maintenance |
| SDS - MIUR | ING-IND/06 | Fluid dynamics |
| | ING-IND/08 | Fluid mechanics |
| | ING-IND/12 | Mechanical and thermal measuring systems |
| | ING-IND/17 | Industrial and mechanical plant |
| | ING-IND/07 | Aerospace propulsion |
| | ING-IND/09 | Energy and environmental systems |

*Table 1: Declared and ascertained competencies for an academic listed in the register of experts.*

| UDA | Total SDSs | No. of experts | Of which academics | SDSs not covered (%) | SDSs covered by a single expert (%) |
|---|---|---|---|---|---|
| Civil engineering | 22 | 130 (6%) | 59 (5%) | 2 (9%) | 1 (5%) |
| Industrial and information engineer. | 42 | 774 (38%) | 509 (42%) | 0 (0%) | 0 (0%) |
| Agriculture and veterinary science | 30 | 147 (7%) | 86 (7%) | 3 (10%) | 4 (13%) |
| Biology | 19 | 207 (10%) | 113 (9%) | 1 (5%) | 1 (5%) |
| Chemistry | 12 | 238 (12%) | 143 (12%) | 0 (0%) | 0 (0%) |
| Earth science | 12 | 54 (3%) | 21 (2%) | 1 (8%) | 2 (17%) |
| Physics | 8 | 196 (10%) | 100 (8%) | 0 (0%) | 0 (0%) |
| Mathematics and computer science | 10 | 176 (9%) | 88 (7%) | 0 (0%) | 1 (10%) |
| Medicine | 50 | 117 (6%) | 89 (7%) | 11 (22%) | 16 (32%) |
| Total | 205 | 1,492 | 956 | 18 (9%) | 25 (12%) |

*Table 2: Number of experts per each UDA of the hard sciences and level of coverage of the SDSs of each UDA.*

| UDA | Academic experts | National academic staff (average 2004-2008) | Concentration index |
|---|---|---|---|
| Civil engineering | 40 (4%) | 1,455 (4%) | 1.14 |
| Industrial and information engineering | 438 (46%) | 5,488 (14%) | 3.30 |
| Agriculture and veterinary science | 74 (8%) | 3,153 (8%) | 0.97 |
| Biology | 80 (8%) | 5,792 (15%) | 0.57 |
| Chemistry | 99 (10%) | 3,610 (9%) | 1.13 |
| Earth science | 13 (1%) | 1,440 (4%) | 0.37 |
| Physics | 78 (8%) | 2,872 (7%) | 1.12 |
| Mathematics and computer science | 49 (5%) | 3,516 (9%) | 0.58 |
| Medicine | 85 (9%) | 12,186 (31%) | 0.29 |
| Total | 956 | 39,512 | |

*Table 3: Number of academic experts and their representativeness of total Italian academic staff, by UDA.*

| UDA | Non-active | IV quart. | III quart. | II quart. | I quart. |
|---|---|---|---|---|---|
| Civil engineering | 17 (43%) | 9 (23%) | 2 (5%) | 4 (10%) | 8 (20%) |
| Industrial and information engineering | 9 (2%) | 121 (28%) | 66 (15%) | 115 (26%) | 127 (29%) |
| Agriculture and veterinary science | 10 (14%) | 11 (15%) | 12 (16%) | 19 (26%) | 22 (30%) |
| Biology | 1 (1%) | 5 (6%) | 12 (15%) | 20 (25%) | 42 (53%) |
| Chemistry | 1 (1%) | 13 (13%) | 12 (12%) | 23 (23%) | 50 (51%) |
| Earth science | 0 (0%) | 3 (23%) | 2 (15%) | 3 (23%) | 5 (38%) |
| Physics | 4 (5%) | 12 (15%) | 12 (15%) | 22 (28%) | 28 (36%) |
| Mathematics and computer science | 2 (4%) | 3 (6%) | 5 (10%) | 15 (31%) | 24 (49%) |
| Medicine | 4 (5%) | 13 (15%) | 9 (11%) | 15 (18%) | 44 (52%) |
| Total | 48 (5%) | 190 (20%) | 132 (14%) | 236 (25%) | 350 (37%) |

*Table 4: Bibliometric performance (Fractional Scientific Strength) of academic experts listed in the MIUR register, by UDA; observation period 2004-2008.*

| UDA | Total SDS | No. of SDSs with average performance ≥ 80 percentile | No. of SDSs with average performance ≤ 50 percentile |
|---|---|---|---|
| Civil engineering | 20 | 0 | 3 |
| Industrial and information engineering | 42 | 2 | 21 |
| Agriculture and veterinary science | 27 | 4 | 7 |
| Biology | 18 | 4 | 2 |
| Chemistry | 12 | 3 | 0 |
| Earth science | 11 | 1 | 3 |
| Physics | 8 | 1 | 1 |
| Mathematics and computer science | 10 | 3 | 0 |
| Medicine | 39 | 10 | 8 |
| Total | 187 | 28 | 45 |

*Table 5: Distribution of bibliometric performance (Fractional Scientific Strength) of academic experts indexed in the MIUR register in the SDSs of each UDA.*

| UDA | Observations | | SDS coverage | | | % of applicants from SDS not covered by referees |
|---|---|---|---|---|---|---|
| | Applic. | Referee | Applic. | Referee | Differ. | |
| Civil engineering | 16 | 3 | 6 | 3 | 3 | 31 |
| Industrial and information engine. | 41 | 8 | 23 | 6 | 17 | 83 |
| Agriculture and veterinary science | 43 | 8 | 27 | 6 | 21 | 77 |
| Biology | 65 | 9 | 17 | 6 | 11 | 52 |
| Chemistry | 32 | 8 | 11 | 6 | 5 | 44 |
| Earth science | 14 | 3 | 9 | 3 | 6 | 71 |
| Physics | 17 | 2 | 5 | 2 | 3 | 59 |
| Mathematics and computer science | 6 | 2 | 4 | 2 | 2 | 50 |
| Medicine | 93 | 12 | 28 | 10 | 18 | 52 |
| Total | 327 | 55 | 130 | 44 | 86 | 58 |

*Table 6: Scientific coverage of referees with respect to the fields represented by applicants in the call for proposals, 2008, Region of Sardinia.*

| UDA | Referees | Of which under the national median | Inactive referees |
|---|---|---|---|
| Civil engineering | 3 | 0 (0%) | 0 (0%) |
| Industrial and information engineering | 8 | 4 (50%) | 2 (25%) |
| Agriculture and veterinary science | 8 | 5 (63%) | 3 (38%) |
| Biology | 9 | 4 (44%) | 1 (11%) |
| Chemistry | 8 | 4 (50%) | 1 (13%) |
| Earth science | 3 | 1 (33%) | 1 (33%) |
| Physics | 2 | 0 (0%) | 0 (0%) |
| Mathematics and computer science | 2 | 1 (60%) | 0 (0%) |
| Medicine | 12 | 3 (24%) | 0 (0%) |
| Total | 55 | 22 (40%) | 8 (15%) |

*Table 7: Bibliometric performance (percentile of Fractional Scientific Strength) of referees in the call for proposals, 2008, Region of Sardinia*